\documentclass[12pt]{article}

\usepackage{amssymb, amsmath}
\usepackage{graphicx}
\usepackage{cite}

\def\be{\begin{equation}}
\def\ee{\end{equation}}
\def\bea{\begin{eqnarray}}
\def\eea{\end{eqnarray}}

\title{ \bf{Extremal Sasakian horizons}}

\author{Hari K. Kunduri$^a$\footnote{hkkunduri@mun.ca } \  and James Lucietti$^b$\footnote{j.lucietti@ed.ac.uk } \\ \\
\small \sl $^a$ Department of Mathematics and Statistics, \\  \small \sl Memorial University of Newfoundland \\ \small \sl St John's NL A1C 4P5, Canada
\\ \small \sl $^b$  School of Mathematics and Maxwell Institute of Mathematical Sciences, \\ \small \sl University of Edinburgh, \\ \small \sl   King's Buildings, Edinburgh, EH9 3JZ, UK }

\date{}

\begin{document}

\maketitle

\begin{picture}(0,0)(0,0)
\put(350, 300){EMPG-12-06}
\end{picture}






\begin{abstract}
We point out a simple construction of an infinite class of Einstein near-horizon geometries in all odd dimensions greater than five. Cross-sections of the horizons are inhomogeneous Sasakian metrics (but not Einstein) on $S^3\times S^2$ and more generally on Lens space bundles over any compact positive K\"ahler-Einstein manifold. They are all consistent with the known  topology and symmetry constraints for asymptotically flat or globally AdS black holes.
\end{abstract}

\vspace{1cm}

\noindent{ {\it Introduction}:   The classification of stationary black hole solutions in higher dimensional General Relativity is an important open problem~\cite{ERrev}, relevant to modern studies in quantum gravity. Unlike in four dimensions, black hole uniqueness is violated even for asymptotically flat vacuum spacetimes~\cite{ER1}, however certain general results are known. Most notably the horizon topology is known to be positive Yamabe type~\cite{GS} and the rigidity theorem guarantees that a rotating black hole has at least one commuting rotational isometry~\cite{HIW, IM, HIRig}. It is not known whether these conditions are sufficient for existence of a black hole solution:  in fact very few explicit examples are known~\cite{MP, ER1, PS} (there are more solutions with disconnected horizons).

Extremal black holes are of particular importance in quantum gravity due to the fact they do not radiate. Hence a classification of such objects is particularly sought after. In fact it turns out that extremal black holes can be partly constrained by their near-horizon geometries. These can be studied independently of a full black hole solution thus providing a simplified setup in which to investigate issues such as horizon topology and symmetry of solutions~\cite{Reall, KLR}. For simplicity we will focus our discussion on the vacuum Einstein equations (possibly allowing for a cosmological constant). In four and five dimensions near-horizon geometries with an appropriate number of commuting rotational symmetries have been classified~\cite{Haj, LP, KL1, KL2}.  In higher dimensions this has not yet been achieved, although some partial results are known~\cite{CRT, FKLR, HI}. In even dimensions certain classes of  near-horizon solutions have been found which are different to the known (spherical) black hole solutions, but possess topology and symmetry consistent with the known general constraints for asymptotically flat and globally AdS black hole solutions~\cite{KL3}. It is the purpose of this note to present new examples, with non-trivial horizon topology, in all odd dimensions, again consistent with the topology and symmetry constraints for asymptotically flat and globally AdS black holes.

Motivated by the above considerations, the precise setup we will consider is as follows. Consider a $D$ dimensional spacetime containing a (smooth) degenerate Killing horizon $\mathcal{N}$ of a Killing vector field $n$, with a  cross section $H$. In the neighbourhood of such a horizon, the spacetime metric written can be written in Gaussian null coordinates (see e.g.~\cite{FRW})
\be
ds^2= 2dv \left( dr + r\, h_A(r,x) dx^A + \frac{1}{2} r^2 F(r,x)dv \right) + g_{AB}(r,x) dx^A dx^B
\ee
where $n=\partial / \partial v$. The vector field $l=\partial / \partial r$ is tangent to the unique null geodesics that start on $\mathcal{N}$ which satisfy $n \cdot l =1$ and are orthogonal to tangent vectors on $H$. The coordinate $r$ is an affine parameter and chosen such that $r=0$ corresponds to $\mathcal{N}$, whereas $(x^A)$ with $A=1, \dots, D-2$, are coordinates on a cross section $H$ of the horizon. We assume that $H$ is a compact manifold (without a boundary). It is well known that for such spacetimes the Einstein equations restrict to a set of equations on $H$ depending only on intrinsic data.
We will consider vacuum spacetimes and allow for a cosmological constant $R_{\mu\nu}=\Lambda g_{\mu\nu}$, with particular interest in $\Lambda\leq 0$. In this case one can show that (see e.g.~\cite{CRT,KLR})
\bea
\label{Heq}
R_{AB} &=& \frac{1}{2}h_Ah_B -\nabla_{(A} h_{B)} +\Lambda g_{AB}  \\  \label{Feq}
F &=& \frac{1}{2}h_A h^A -\frac{1}{2}\nabla_A h^A +\Lambda
\eea
where $R_{AB}$ and $\nabla_A$ are the Ricci tensor and metric connection associated to the induced (Riemannian) metric on $H$ defined by $g_{AB}\equiv g_{AB}(0,x)$, and $h_a\equiv  h_a(0,x)$ and $F \equiv F(0,x)$.  One can understand this result in terms of the so-called near-horizon limit. This is defined by replacing $r \to \epsilon r$ and $v \to v/ \epsilon$ and then taking the limit $\epsilon \to 0$~\cite{Reall,CRT,KLR}. The limit always exists for a (smooth) degenerate horizon resulting in the {\it near-horizon geometry}
\be
ds^2_{NH} = 2dv \left( dr + r h_A(x) dx^A+ \frac{1}{2} r^2 F(x)dv \right) + g_{AB}(x) dx^A dx^B  \; .
\ee
One can then check that $ds^2_{NH}$ satisfies $R_{\mu\nu} = \Lambda g_{\mu\nu}$ if and only if (\ref{Heq}) and (\ref{Feq}) are satisfied. In this note we will present a simple set of solutions to the near-horizon equations (\ref{Heq}) in all odd dimensions.

\vspace{.5cm}

\noindent{ \it Construction of near-horizon geometries}: We will make the following assumption: $h^A$ is a Killing vector field. The horizon equation then reduces to
\be
\label{eta}
R_{AB} = \frac{1}{2}h_Ah_B +\Lambda g_{AB}  \; .
\ee
The contracted Bianchi identity then implies that $|h|^2= h_Ah^A$ is a constant (it follows that $h^A$ is also tangent to geodesics). Therefore the scalar curvature $R=\frac{1}{2}|h|^2+ (D-2)\Lambda$ is a constant, which we assume to be positive. Also note that equation (\ref{Feq}) gives $F=\frac{1}{2}|h|^2 +\Lambda$ (and hence a constant). We deduce that for a non-static solution, $h^A$ is a nowhere vanishing vector field on $H$. This is only possible if the Euler characteristic vanishes $\chi(H)=0$, a constraint which is of course automatically satisfied in all odd dimensions\footnote{In even dimensions $\chi(H)=0$ of course provides a non-trivial constraint on the topology. For example, it implies that $H$ cannot be simply connected.}. It is worth noting that any homogeneous near-horizon geometry necessarily has $h$ Killing\footnote{By a homogeneous near-horizon geometry we mean $(H,\gamma)$ is a homogeneous space such that the near-horizon data $(h, F)$ is invariant under the isometries of $\gamma$. It easily follows that $h^A$ is a Killing field.} so our considerations include such cases, however we will not assume homogeneity.  Now there are two possibilities: either the orbits of $h$ are closed or they are open. In the former case the orbit space is itself a compact manifold, which we refer to as the transverse space $T$. In the latter case the quotient space is only locally defined, but we still refer to it as the transverse space $T$.

It is convenient to use coordinates on $H$ adapted to the Killing field $h^A$. Writing $(x^A)=(\bar{\psi},y^a)$ where $a=1, \dots, D-3$, so $y^a$ are coordinates on the transverse space $T$,  our near-horizon data reads
\be
g_{AB}dx^Adx^B = (d\bar{\psi}+\hat{\sigma}_a dy^a)^2 +\hat{g}_{ab}dy^ady^b \; ,  \qquad \qquad h = k \frac{\partial}{\partial \bar{\psi}}
\ee
where $\hat{\sigma}_a$ and $\hat{g}_{ab}$ are a 1-form and metric on $T$, and $k$ is a constant so $h_Ah^A=k^2$. This parameterisation is useful as $k=0$ then corresponds to an Einstein manifold.  Now define a 2-form on $T$ by $\hat{\omega}=\frac{1}{2}d\hat{\sigma}$. The horizon equation (\ref{Heq}) is then equivalent to a set of equations on the $D-3$ dimensional transverse space $T$:
\be
\label{Teq}
\hat{R}_{ab} = 2\hat{\omega}_{ac} \hat{\omega}_{b}^{~c} +\Lambda \hat{g}_{ab}, \qquad \qquad \hat{\nabla}^a \hat{\omega}_{ab}=0, \qquad  \qquad \hat{\omega}_{ab} \hat{\omega}^{ab}=\frac{k^2}{2}+\Lambda
\ee
where $\hat{R}_{ab}$ and $\hat{\nabla}_a$ are the Ricci curvature and Levi-Civita connection associated to the transverse metric $\hat{g}_{ab}$. We will assume that the two-form $\hat{\omega}$ is non-trivial and hence require that $\frac{k^2}{2}+\Lambda>0$. It is worth noting that the full near-horizon geometry can be written as
\be
ds^2_{NH} = \left( -\frac{k^2}{2}+\Lambda \right) r^2 dv^2 +2dvdr + (d\bar{\psi}+ \hat{\sigma}_a dy^a + k rdv )^2 +\hat{g}_{ab} dy^a dy^b
\ee
which is an $H$-bundle over AdS$_2$ (if $ -\frac{k^2}{2}+\Lambda <0$) preserving the full $SO(2,1)$ symmetry of AdS$_2$, as is typically the case for near-horizon geometries~\cite{KLR, FKLR}. Hence all the solutions we discuss in this note also enjoy this near-horizon symmetry enhancement.

The set of equations (\ref{Teq}) are difficult to solve in general\footnote{Indeed a similar set of equations arise in the classification of homogeneous metrics on circle bundles over K\"ahler-Einstein manifolds~\cite{Besse}.}. However, let us now make a further assumption. Namely that the transverse space is $2n$-dimensional with K\"ahler structure $(T, \hat{g}, \hat{J})$ and the K\"ahler form $\hat{J}$ proportional to $\hat{\omega}$. This is equivalent to requiring our horizon to be locally Sasakian with $h^A$ proportional to the Reeb vector. The horizon equations now simplify to
\be \label{transverseeq}
\hat{\omega}_{ab} = \sqrt{ \frac{\frac{k^2}{2}  +\Lambda}{2n} }  \hat{J}_{ab} \; , \qquad \qquad
\hat{R}_{ab} = \left( \frac{k^2 +2\Lambda(n+1)}{2n} \right) \hat{g}_{ab}
\ee
and thus the transverse space is also Einstein. It is worth pausing here to point out that for $n=1$ this is the only solution to (\ref{Teq}) and hence $\hat{g}$ is locally isometric to the round $S^2$. The resulting horizon geometry is then a homogeneous metric on $S^3$ (or quotients) and the near-horizon geometry is locally isometric to that of the extremal Myers-Perry(-AdS) black hole with equal angular momenta. This therefore solves the classification problem for $D=5$ homogeneous Einstein near-horizon geometries. From now on we will assume $n \geq 2$ so $D=2n+3\geq 7$.

So far we have shown that given any K\"ahler-Einstein metric we can construct a solution to the horizon equation which is (locally) Sasakian. If $T$ is a manifold then we may take any compact positive K\"ahler-Einstein manifold. For example, $T = \mathbb{CP}^n$ with the Fubini-Study metric gives a homogeneous horizon metric on $S^{2n+1}$, which corresponds to that of the extremal Myers-Perry black hole in odd dimensions with all angular momenta equal (see e.g.~\cite{FKLR}). One could also choose $T=\mathbb{CP}^1\times \mathbb{CP}^1$ which yields a homogeneous horizon metric on $T^{1,1} \cong S^3 \times S^2$. On the other hand, one may also choose a K\"ahler-Einstein base with no continuous isometries such as the higher del Pezzo surfaces, which results in a horizon geometry with a single rotational symmetry. These would saturate the lower bound for the rigidity theorem for black holes and are odd dimensional counterparts of the solutions in~\cite{KL3}.

We may construct simple non-spherical horizon topologies by following the construction of the Sasaki-Einstein manifolds~\cite{GMSW1, GMSW}. Thus, we use the following K\"ahler metric on $T$:
\bea \label{Kahlerbase}
\hat{g}_{ab}dy^ady^b &=& \frac{x^{n-1} dx^2}{2P(x)} + \frac{2P(x)}{x^{n-1}}( d\bar{\phi}+\bar{\sigma})^2 + 2x \bar{g}   \\
\hat{J}&=&d[ x( d\bar{\phi}+\bar{\sigma} )]   \label{Kahlerform}
\eea
where $\text{Ric}(\bar{g})=2n \bar{g}$ and $\bar{g}$ is itself a $2n-2$ dimensional K\"ahler-Einstein metric on a base $K$ with K\"ahler form $\bar{J}=\frac{1}{2} d\bar{\sigma}$ . For $n=2$ the only choice is $K =\mathbb{CP}^1 \cong S^2$ with round metric, which gives the cohomogeneity-1 K\"ahler metric with $SU(2)\times U(1)$ symmetry~\cite{PP}. The Einstein condition $\text{Ric}(\hat{g})=\hat{\lambda} \hat{g}$ is then simply:
\be
P(x)=x^n - \frac{\hat{\lambda} x^{n+1}}{n+1} +c
\ee
where $c$ is a constant.
Therefore in order to get a solution to our horizon equation we simply require
\be \label{KEconstant}
\hat{\lambda}= \frac{k^2 +2\Lambda(n+1)}{2n}  \; .
\ee
To summarise our horizon metric takes the form
\be \label{etaSasaki}
g_{AB}dx^Adx^B= [d\bar{\psi}+2\omega_0x(d\bar{\phi}+\bar{\sigma})]^2 + \frac{x^{n-1} dx^2}{2P(x)} + \frac{2P(x)}{x^{n-1}}( d\bar{\phi}+\bar{\sigma})^2 + 2x \bar{g}  
\ee
where for convenience we have defined the positive constant
\be
\omega_0 \equiv  \sqrt{ \frac{\frac{k^2}{2}  +\Lambda}{2n} }  \;.
\ee
The Sasaki-Einstein case is recovered by simply setting $k=0$ (and furthermore one has to have $\Lambda=2n>0$). Thus we see that we have 1-parameter  deformation of the Sasaki-Einstein spaces, given by the parameter $k$, which is contained entirely in the constant $\hat{\lambda}$. These deformations preserve the (local) Sasakian structure but are of course not Einstein.  Crucially, we may have $\Lambda\leq 0$, as required for their interpretation as horizons of asymptotically flat or AdS black holes.

In fact the horizon equation (\ref{eta}) is sometimes called the ``$\eta$-Einstein" condition and so the solutions we have obtained are examples of Sasaki $\eta$-Einstein manifolds~\cite{eta}.\footnote{It is also worth emphasising that they have constant scalar curvature, and therefore amusingly they are also extremal in the sense used in geometry (i.e. they are critical points of $\int R^2$)~\cite{Boyer}.} By convention, a Sasaki $\eta$-Einstein manifold of dimension $2n+1$ is normalised so that  $\omega_0=1$. We will not fix this normalisation, but still refer to our class of horizon solutions as Sasakian, as one merely needs to rescale the metric by a constant factor to recover the standard definition.

\vspace{.5cm}

\noindent{ {\it Global analysis}: We now turn to a global analysis of the above metric using the same strategy as in the Einstein case~\cite{GMSW1, GMSW} so we will be brief. We assume $\hat{\lambda}>0$ which allows for $\Lambda \leq 0$ when $k \neq 0$.  We want to find conditions such that our local metric (\ref{etaSasaki}) extends to a smooth metric on a compact manifold $H$. The local metric has potential singularities at $x=0$ and the roots of $P(x)$. Compactness requires that we take $x_1 \leq x \leq x_2$ with $P(x)\geq 0$, where $x_1,x_2$ are two adjacent positive simple roots of $P(x)$. The only turning points of $P(x)$ are at $x=0, n/\hat{\lambda}$ and thus its graph looks like a negative cubic. The necessary and sufficient conditions for existence of roots $0<x_1<x_2$ are:
\be
\label{clambdaineq}
c_* \equiv -\frac{1}{n+1} \left( \frac{n}{\hat{\lambda}} \right)^n <c <0  \; .
\ee
We now make a convenient local change of coordinates $(\bar{\psi}, \bar{\phi}) \to (\psi, \phi)$ defined by $\bar{\psi}=- 2n \omega_0 \psi/ \hat{\lambda}$ and $\bar{\phi} = \phi/n  +\psi$. The horizon metric can then be written as a local $U(1)$ fibration over a different base space $B$:
\be
\label{U1bundle}
g = A(x) [ d\psi + A_B ]^2  + g_B \; , \qquad \qquad A_B = \Omega(x) (d\phi+ n \bar{\sigma})
\ee
where    $A(x) = \frac{2P(x)}{x^{n-1}} + 4\omega_0^2( x- \frac{n}{\hat{\lambda}})^2$ and $\Omega(x) =\frac{1}{n A(x)}\left[  \frac{2P(x)}{x^{n-1}} + 4\omega_0^2x( x- \frac{n}{\hat{\lambda}}) \right]$ and
\be
\label{base}
g_B = \frac{x^{n-1} dx^2}{2P(x)} + \frac{8\omega_0^2 P(x)}{\hat{\lambda}^2 A(x) x^{n-1}}( d\phi+n\bar{\sigma})^2 + 2x \bar{g}   \; .
\ee
Now, using the explicit form for $P(x)$, it is easily seen that  at any root $4\omega_0^2 P'(x_i)^2 = \hat{\lambda}^2 A(x_i) x_i^{2n-2}$. It can be checked that this implies one can simultaneously remove the conical singularities at $x=x_i$ in the $(x,\phi)$ part of the metric $g_B$. In particular, if we choose the coordinate $\phi$ to be periodic with period $\Delta \phi =2\pi$, the $(x, \phi)$ part of the base metric $g_B$ extends to a smooth metric on $S^2$. We will make this choice, so that $g_B$ gives a metric which is locally an $S^2$-bundle over the K\"ahler-Einstein base $(K, \bar{g})$. In order for this bundle to be globally defined on a compact total space $H$, one needs that the associated $U(1)$-bundle is regular and $K$ is a compact manifold. The normalised connection for this $U(1)$-bundle is $A=d\phi+n \bar{\sigma}$ and so the periods for a basis of 2-cycles $\Sigma_I \subset K$ are
\be
\frac{1}{2\pi} \int_{\Sigma_I} dA=  \frac{n}{2\pi} \int_{\Sigma_I} 2\bar{J} = \int_{\Sigma_I} c_1(K) \equiv c_I
\ee
where we have used $\bar{\rho}= 2n \bar{J}$, where $\bar{\rho}$ is the Ricci form of $K$, and that the first Chern class $c_1(K)= [ \bar{\rho} /2\pi ]$. Because the latter is an integral class these periods $c_I \in \mathbb{Z}$ (Chern numbers) and hence the bundle is automatically regular. Furthermore since $c_1(K)= -c_1(\mathcal{L})$, where $\mathcal{L}$ is the canonical line bundle, we see that the $U(1)$ bundle is isomorphic to $\mathcal{L}^{*}$. To summarise, we have so far shown that $g_B$ extends to a smooth metric on a base manifold $B$, which is an $S^2$-bundle over $K$, if and only if $\Delta \phi =2\pi$. Furthermore, this bundle is in fact the associated $S^2$-bundle to the anti-canonical line bundle over $K$. For $n=2$ we have $K \cong S^2$ and $c_I$ are always even; hence $B$ is a trivial $S^2$-bundle over $S^2$, so $B \cong S^2\times S^2$.

The final part of the regularity analysis involves showing that our horizon metric $g$ can be extended to a smooth metric on a $U(1)$ bundle over $B$, for a countable choice of values of $c$ in the range (\ref{clambdaineq}). The 1-form $A_B$ may be thought of as a $U(1)$ connection on $B$ and it is readily verified that its curvature $dA_B$ is globally defined on $B$. If we let the period of $\psi$ be given by $\Delta \psi = 2\pi \ell$, then regularity of the $U(1)$-bundle over $B$ defined by this connection requires that the periods of $\frac{dA_B}{2\pi \ell}$ over a basis of 2-cycles for $B$ are integers. A basis of 2-cycles for $B$ is given by the $S^2$ fibre $\Sigma$ (at a fixed base point on $K$) and by the two cycles $s\Sigma_I$, where the section $s: K \to B$ maps to the pole $x=x_2$ (say). Then we compute the Chern numbers
\bea
 \frac{1}{2\pi\ell} \int_\Sigma dA_B  &=& \frac{\Omega(x_2)-\Omega(x_1)}{\ell} \\
 \frac{1}{2\pi \ell} \int_{s\Sigma_I} dA_B &=&  \frac{\Omega(x_2)}{\ell} \int_{\Sigma_I} c_1(K)  = \frac{\Omega(x_2) c_I}{\ell}
\eea
where $c_I$ are the Chern numbers as above. We deduce that
\be
\label{pqdef}
\frac{\Omega(x_2)-\Omega(x_1)}{\ell} =p, \qquad \qquad   \frac{\Omega(x_2)}{\ell}= \frac{q}{I}
\ee
where $p,q$ are non-zero integers and $I \equiv \text{gcd}(c_I)$ (the Fano index of $K$). It follows that
\be
\label{quant}
\frac{\Omega(x_1)}{\Omega(x_2)} = 1- \frac{Ip}{q} \; .
\ee
Hence existence of regular solutions reduces to solving the equation (\ref{quant}). Explicitly
\be
R(c; \hat{\lambda}) \equiv \frac{\Omega(x_1)}{\Omega(x_2)} = \frac{ x_1 \left( x_2 -\frac{n}{\hat{\lambda}} \right) }{x_2 \left( x_1 -\frac{n}{\hat{\lambda}} \right)}
\ee
defines a continuous function of $c$ (for each $\hat{\lambda}>0$), in the range (\ref{clambdaineq}). As $c \to 0^-$ we have $x_1 \to 0^+, x_2 \to (n+1)^-/ \hat{\lambda}$ and $R(c; \hat{\lambda}) \to 0^- $. As $c \to (c_*)^+$ we have $x_1 \to (n/\hat{\lambda})^-,  x_2 \to (n/\hat{\lambda})^+$ and $R(c; \hat{\lambda}) \to (-1)^+$. Therefore, for each $\hat{\lambda}>0$, $R(c; \hat{\lambda})$ is a continuous function of $c \in (c_* , 0)$ such that in the limit $R(0, \hat{\lambda})=0$ and $R(c_*, \hat{\lambda})=-1$. It immediately follows that for every pair of integers $(p,q)$ such that $1<Ip/q<2$, there exist a solution to (\ref{quant}).

To summarise, we have shown that there exists a countably infinite number of smooth compact horizons (for each value of the continuous parameter $k$), labelled by pairs of integers $(p,q)$ satisfying $1<Ip/q<2$. For $k=0$ these reduce to the known Sasaki-Einstein manifolds~\cite{GMSW1,GMSW}, whereas for $k \neq 0$ these give $\eta$-Einstein Sasakian manifolds of a similar nature (it is readily checked the Sasakian structure is globally defined). We will call these horizon manifolds $H^{p,q}$.

\vspace{.5cm}

\noindent{{\it Topology}}: As we have seen $H^{p,q}$ is the total space of a $U(1)$-bundle over $B$ with Chern numbers $(p,q)$. The topology of these spaces is the same as the Sasaki-Einstein manifolds of~\cite{GMSW1, GMSW}, so again we will be brief. The base manifold $B$ is itself an $S^2$-bundle over a compact positive K\"ahler-Einstein manifold $K$. As $K$ must be simply connected, it follows that $B$ is as well. Furthermore, if $p,q$ are co-prime, it follows that $H^{p,q}$ is also simply connected, which we assume henceforth. One can also show that $B$ and $H^{p,q}$ are spin manifolds. For $n=2$ we showed earlier $B \cong S^2\times S^2$ and one can show that $H^{p,q} \cong S^3\times S^2$. More generally, $H^{p,q}$ is a Lens space bundle over $K$. This can be seen as follows.

At a fixed base point $K$ the 3d fibre is compact and has a $U(1)^2$ isometry generated by the $2\pi$-normalised Killing fields $(\ell \partial_\psi, \partial_\phi)$. For each $i=1,2$, the canonically normalised Killing vector field $K_i= \Omega(x_i)\partial_\psi- \partial_\phi$ vanishes on exactly one codimension-2 submanifold given by $x=x_i$. These two sets of Killing fields are related by the matrix
\be
\left( \begin{array}{c} K_1 \\ K_2 \end{array} \right) = \left( \begin{array}{cc} \frac{q}{I}-p & -1 \\ \frac{q}{I} & -1 \end{array} \right)\left( \begin{array}{c} \ell \partial_\psi \\ \partial_\phi \end{array} \right)\; ,
\ee
where we have used (\ref{pqdef}). The determinant of this matrix is $p$, which allows us to deduce that the fibre is a Lens space $S^3/\mathbb{Z}_p \cong L(p,1)$.

\vspace{.5cm}

\noindent{{\it Physical quantities}}: The area of the horizons $H^{p,q}$ is
\be
A(H^{p,q}) = \frac{2^{n+2} \pi^2  \ell \omega_0\, (x_2^n-x_1^n) \;\text{vol}(K)}{n \hat{\lambda}}
\ee
where $\text{vol}(K)$ is the volume of $K$.
The Komar angular momentum with respect to a rotational Killing field $m$, for a spacetime containing a degenerate Killing horizon, can be evaluated as an integral over the horizon~\cite{FKLR}
\be
j[m] = \frac{1}{16 \pi} \int_H  \sqrt{g} \; h \cdot m  \; .
\ee
For our horizons $H^{p,q}$ we get
\bea
j[\partial_\psi] &=& 0, \\  j[\partial_\phi] &=& \frac{2^{n-1} \pi \ell \omega_0^2  k (x_2^{n+1}-x_1^{n+1}) \text{vol}{(K)} }{\hat{\lambda}  n(n+1)}, \\  j[\bar{m}] &=&  \frac{\int_K \bar{\sigma} \cdot \bar{m}}{\text{vol}(K)}  \; j[\partial_\phi]
\eea
where $\bar{m}$ is a rotational Killing field on $K$ (should any exist). Somewhat surprisingly, we see there is no angular momentum in the direction of the $U(1)$-fibre ($\psi$). In~\cite{KL3} it was shown that $\int_K \bar{\sigma} \cdot \bar{m}=0$ for toric $K$, so in this case the spins associated to $K$ also vanish.

\vspace{.5cm}

\noindent{{\it Generalisations:}}  Our construction admits a straightforward generalisation that gives  horizon solutions with the same topology as the Sasaki-Einstein manifolds $L^{p,q,r}$, which in five dimensions are all diffeomorphic to $S^3 \times S^2$~\cite{CPPL, MS}. To generalise our five dimensional horizons, we need only replace (\ref{Kahlerbase}, \ref{Kahlerform}) with the following toric  K\"ahler metric \cite{ACG, MS}\footnote{This in fact represents the most general orthotoric K\"ahler surface~\cite{ACG}.}
\bea
\hat{g}_{ab}dy^ady^b &=& \frac{x-y}{2Q(y)} dy^2  + \frac{2Q(y)}{x-y} (d\hat{\phi}+ xd\hat{\psi})^2+ \frac{x-y}{2P(x)} dx^2 + \frac{2P(x)}{x-y}(d\hat{\phi}+y d\hat{\psi})^2 \label{toricKB}\\ 
\hat{J} &=& d[ (x+y)d\hat{\phi} +xy d\hat{\psi}]  \label{toricJ}
\eea
for which the Einstein condition $\text{Ric}(\hat{g})= \hat{\lambda} \hat{g}$ is
\be
Q(y)= \frac{\hat{\lambda}}{3}y(\alpha-y)(\alpha-\beta -y) \; , \qquad \quad P(x)= -\frac{\hat{\lambda}}{3} x(\alpha-x)(\alpha-\beta-x) +c
\ee
where $\alpha, \beta, c$ are integration constants and without loss of generality we have used the translation freedom $(x,y, \hat{\phi}, \hat{\psi})\mapsto (x+c, y+c, \hat{\phi}-c\hat{\psi}, \hat{\psi})$ to fix one of the roots of $Q(y)$ to zero.
By a change of coordinates this base may be written in such a way that it contains the $n=2$ cohomogeneity-1 metric (\ref{Kahlerbase}). This is given by first assuming $\alpha \neq \beta$ and setting $y = (\alpha-\beta) \sin^2\left(\frac{\theta}{2}\right)$, $\hat{\phi} \to \hat{\psi}/2$ and $\hat{\psi} \to (\beta\hat{\phi} -\alpha \hat{\psi})/[2\alpha(\alpha-\beta)]$. The base in the resulting coordinates then in fact allows one to set $\alpha=\beta=3/\hat{\lambda}$ (the last equality is without loss of generality), which then reduces precisely to the cohomogeneity-1 case (\ref{Kahlerbase}, \ref{Kahlerform}) with $\bar{\phi} = (\hat{\phi} + \hat{\psi})/4$, $\bar{\sigma}= (1/2) \cos \theta d\bar{\chi}$ and $\bar{g}=(1/4)( d\theta^2 +\sin^2\theta d\bar{\chi}^2)$, where $\bar{\chi}=(\hat{\psi}-\hat{\phi})/2$. The global analysis of the resulting horizon manifold for $\alpha \neq \beta$ can be performed as in the Einstein case $L^{p,q,r}$ \cite{CPPL,MS}, so we omit details. A similar construction can be performed to find cohomogeneity-$n$ horizons with dimension $2n+1$ with the same topology as the spaces $L^{p,q,r_1\cdots r_{n-1}}$ obtained in \cite{CPPL}.

\vspace{.5cm}

\noindent{{\it Summary:}} To summarise, we have given a simple construction of an infinite class of vacuum near-horizon geometries, allowing for a cosmological constant, in all odd dimensions $D=2n+3$ greater than five. The horizon geometries are inhomogeneous Sasakian metrics on $S^3\times S^2$ or more generally on $L(p,1)$-bundles over any compact positive K\"ahler-Einstein manifold $K$. These geometries have isometry group $G \cong U(1)^2 \times G_K$ where $G_K$ is the isometry group of $K$. As a result, the Cartan subgroup of $G$ is a subgroup of $U(1)^{n+1}$ (which is the Cartan subgroup of $SO(2n+2)$), as required for the horizon of an asymptotically flat or globally AdS black hole (for example with $K =  \mathbb{CP}^{n-1}$ it is $U(1)^{n+1}$). Further, the horizon topologies are allowed by the known constraints for such black holes (i.e. positive Yamabe and cobordant to spheres). For a given $K$, they depend on two integers and a continuous parameter which corresponds to the Komar angular momentum of the horizon.  It would of course be interesting to determine whether these are realised as the horizon geometries of yet to be found extremal black holes.

We have also found more general (non-Sasakian) horizon metrics in odd dimensions which include the examples given here. These will be presented elsewhere~\cite{KL4}.

\vspace{.5cm}
\noindent{{\it Acknowledgements}}:  HK is supported by an NSERC Discovery Grant.  JL is supported by an EPSRC Career Acceleration Fellowship. JL would like to thank Paul de Medeiros for useful discussions.


\begin{thebibliography}{99}

{ \small

\bibitem{ERrev}
  R.~Emparan and H.~S.~Reall,
  ``Black Holes in Higher Dimensions,''
  Living Rev.\ Rel.\  {\bf 11} (2008) 6
  [arXiv:0801.3471 [hep-th]].

  \bibitem{ER1}
  R.~Emparan and H.~S.~Reall,
  ``A rotating black ring in five dimensions,''
  Phys.\ Rev.\ Lett.\  {\bf 88} (2002) 101101
  [arXiv:hep-th/0110260].

  \bibitem{GS}
  G.~J.~Galloway and R.~Schoen,
  ``A generalization of Hawking's black hole topology theorem to higher
  dimensions,''
  Commun.\ Math.\ Phys.\  {\bf 266} (2006) 571
  [arXiv:gr-qc/0509107].

  \bibitem{HIW}
S.~Hollands, A.~Ishibashi and R.~M.~Wald,
``A Higher Dimensional Stationary Rotating Black Hole Must be Axisymmetric,'' Commun.\ Math.\ Phys.\  {\bf 271} (2007) 699 [arXiv:gr-qc/0605106].

\bibitem{IM}
  V.~Moncrief and J.~Isenberg,
  ``Symmetries of Higher Dimensional Black Holes,''
  Class.\ Quant.\ Grav.\  {\bf 25}, 195015 (2008)
  [arXiv:0805.1451 [gr-qc]].

  \bibitem{HIRig}
  S.~Hollands and A.~Ishibashi,
  ``On the `Stationary Implies Axisymmetric' Theorem for Extremal Black Holes
  in Higher Dimensions,''
  Commun.\ Math.\ Phys.\  {\bf 291} (2009) 403
  [arXiv:0809.2659 [gr-qc]].

 \bibitem{MP}
  R.~C.~Myers and M.~J.~Perry,
  ``Black Holes In Higher Dimensional Space-Times,''
  Annals Phys.\  {\bf 172} (1986) 304.

  \bibitem{PS}
  A.~A.~Pomeransky and R.~A.~Sen'kov,
  ``Black ring with two angular momenta,''
  arXiv:hep-th/0612005.

\bibitem{Reall}
  H.~S.~Reall,
  ``Higher dimensional black holes and supersymmetry,''
  Phys.\ Rev.\  D {\bf 68} (2003) 024024
  [Erratum-ibid.\  D {\bf 70} (2004) 089902]
  [arXiv:hep-th/0211290].

  \bibitem{KLR}
  H.~K.~Kunduri, J.~Lucietti and H.~S.~Reall,
  ``Near-horizon symmetries of extremal black holes,''
  Class.\ Quant.\ Grav.\  {\bf 24} (2007) 4169
  [arXiv:0705.4214 [hep-th]].

  \bibitem{Haj}
P.~H{a}j{{i}}{{c}}ek, ``Three remarks on axisymmetric stationary horizons'', Commun.Math.\ Phys. {\bf 36} (1974), p.~305--320.

\bibitem{LP}
  J.~Lewandowski and T.~Pawlowski,
  ``Extremal Isolated Horizons: A Local Uniqueness Theorem,''
  Class.\ Quant.\ Grav.\  {\bf 20} (2003) 587
  [arXiv:gr-qc/0208032].

\bibitem{KL1}
  H.~K.~Kunduri and J.~Lucietti,
  ``A classification of near-horizon geometries of extremal vacuum black
  holes,''
  J.\ Math.\ Phys.\  {\bf 50} (2009) 082502
  [arXiv:0806.2051 [hep-th]].
\bibitem{KL2}
  H.~K.~Kunduri and J.~Lucietti,
  ``Uniqueness of near-horizon geometries of rotating extremal AdS(4) black
  holes,''
  Class.\ Quant.\ Grav.\  {\bf 26} (2009) 055019
  [arXiv:0812.1576 [hep-th]].

  \bibitem{CRT}
  P.~T.~Chrusciel, H.~S.~Reall and P.~Tod,
  ``On non-existence of static vacuum black holes with degenerate  components
  of the event horizon,''
  Class.\ Quant.\ Grav.\  {\bf 23} (2006) 549
  [arXiv:gr-qc/0512041].


   \bibitem{FKLR}
  P.~Figueras, H.~K.~Kunduri, J.~Lucietti and M.~Rangamani,
  ``Extremal vacuum black holes in higher dimensions,''
  Phys.\ Rev.\  D {\bf 78} (2008) 044042
  [arXiv:0803.2998 [hep-th]].

  \bibitem{HI}
  S.~Hollands and A.~Ishibashi,
  ``All vacuum near horizon geometries in  $D$-dimensions with $(D-3)$ Commuting Rotational Symmetries,''
  arXiv:0909.3462 [gr-qc].

  \bibitem{KL3}
  H.~K.~Kunduri and J.~Lucietti,
  ``An infinite class of extremal horizons in higher dimensions,''
  Commun.\ Math.\ Phys.\  {\bf 303} (2011) 31
  [arXiv:1002.4656 [hep-th]].

\bibitem{FRW}
  H.~Friedrich, I.~Racz and R.~M.~Wald,
  ``On the rigidity theorem for space-times with a stationary event horizon or a compact Cauchy horizon,''
  Commun.\ Math.\ Phys.\  {\bf 204} (1999) 691
  [gr-qc/9811021].

 \bibitem{Besse}
 A.L. Besse, \emph{Einstein Manifolds}, Springer-Verlag, 2nd edition, 1987.

 \bibitem{eta}
 Charles P. Boyer, Krzysztof Galicki, Paola Matzeu, \emph{On eta-einstein sasakian geometry}, Commun.Math.Phys.262:177-208,2006, [arXiv:math/0406627v4 [math.DG]]
 
 \bibitem{Boyer}
 Charles P. Boyer, \emph{Extremal Sasakian Metrics on $S^3$-bundles over $S^2$}, Mathematical Research Letters 18 (2011), no. 01, 181-189, 	[arXiv:1002.1049v3 [math.DG]]

\bibitem{GMSW1}
  J.~P.~Gauntlett, D.~Martelli, J.~Sparks and D.~Waldram,
  ``Sasaki-Einstein metrics on S(2) x S(3),''
  Adv.\ Theor.\ Math.\ Phys.\  {\bf 8} (2004) 711
  [arXiv:hep-th/0403002].

  \bibitem{GMSW}
  J.~P.~Gauntlett, D.~Martelli, J.~F.~Sparks and D.~Waldram,
  ``A new infinite class of Sasaki-Einstein manifolds,''
  Adv.\ Theor.\ Math.\ Phys.\  {\bf 8} (2006) 987
  [arXiv:hep-th/0403038].

  \bibitem{PP}
  D.~N.~Page and C.~N.~Pope,
  ``Inhomogeneous Einstein Metrics On Complex Line Bundles,''
  Class.\ Quant.\ Grav.\  {\bf 4} (1987) 213.

  \bibitem{CPPL}
  M.~Cvetic, H.~Lu, D.~N.~Page and C.~N.~Pope,
  ``New Einstein-Sasaki and Einstein spaces from Kerr-de Sitter,''
  JHEP {\bf 0907} (2009) 082
  [hep-th/0505223].

 \bibitem{MS}
  D.~Martelli and J.~Sparks,
  ``Toric Sasaki-Einstein metrics on S**2 x S**3,''
  Phys.\ Lett.\ B {\bf 621} (2005) 208
  [hep-th/0505027].
  
 \bibitem{ACG}
 V.~Apostolov, D.~M~.~J.~Calderbank, P.~Gauduchon
 ``The geometry of weakly selfdual Kahler surfaces", Compositio Math. 135 (2003) 279-322, 	[arXiv:math/0104233v2 [math.DG]]

\bibitem{KL4}
H.~K.~Kunduri and J.~Lucietti,
  ``Degenerate horizon, Einstein metrics, and Lens space bundles'',  arXiv:1210.1268 [hep-th].





}

\end{thebibliography}
\end{document}